\documentclass{article}
\usepackage{nips2004e}
\usepackage{epsfig}
\usepackage{amsfonts}
\usepackage{amsmath}
\usepackage{amssymb}
\usepackage{comment}
\usepackage{url}

\newcommand{\jndm}{\mbox{JNDM}~}
\newcommand{\jndmp}{\mbox{JNDM}}
\renewcommand{\vec}[1]{\mathbf{#1}}

\title{Journal of New Democratic Methods\\
An Introduction}

\author{
John David Funge \\
\url{www.democraticjournals.org} \\
\url{jfunge@gmail.com} \\
}

\begin{document}

\maketitle

\begin{abstract}

This paper describes a new breed of academic journals that use
statistical machine learning techniques to make them more democratic.
In particular, not only can anyone submit an article, but anyone can
also become a reviewer.  Machine learning is used to decide which
reviewers accurately represent the views of the journal's readers and
thus deserve to have their opinions carry more weight.  The paper
concentrates on describing a specific experimental prototype of a
democratic journal called the \emph{Journal of New Democratic Methods}
(\jndmp).  The paper also mentions the wider implications that machine
learning and the techniques used in the \jndm may have for
representative democracy in general.\\

\end{abstract}

\section{Introduction}

Anyone can submit an article to an academic journal, but you can
usually only become a reviewer if you are already well respected, or
well connected, within the academic community.  This paper describes a
new kind of academic journal in which anyone can review articles
submitted to the journal.  Machine learning is then used to learn
which reviewers accurately represent the views of the journal's
readers and thus deserve to have their opinions carry more weight.


This paper concentrates on describing a specific experimental
prototype of a democratic journal called the \emph{Journal of New
Democratic Methods} (\jndmp).  When a submitted article is reviewed
for the \jndmp, the past performance of reviewers is used to predict
the probability that, should it be published, the article will be
acceptable to a majority of the article's subsequent readers.
Articles that are determined to have a probability of greater than a
half of being acceptable are published.

Before formally defining what it means for an article to be
acceptable, some definitions are required.  In particular, suppose an
article has $m$ reviewers, then the reviews consist of a vector of
binary values $\vec{r} = (r_1, \ldots, r_m)$ such that $r_i = 1$ if
review $i$ choose to accept the article and $r_i = 0$ otherwise.
Similarly, if an article has $n$ readers who express an opinion, then
the opinions consist of a vector of binary values $\vec{e} = (e_1,
\ldots, e_n)$ such that $e_j = 1$ if reader $j$ thought the article
was acceptable and $e_j = 0$ otherwise.  A published article is said
to be \emph{acceptable} if a simple majority of the article's readers,
who express an opinion, think it is acceptable,
\[
\text{acceptable} \equiv \sum_{j=1}^n e_j > \frac{n}{2}.
\]

Using the previous definitions, the task of deciding whether to accept
an article is formally expressed as determining whether the
conditional probability, given the reviews, that an article will be
acceptable is greater than a half,
\[
P( \text{acceptable}~|~\vec{r} ) > \frac{1}{2}.
\]
To make the acceptance criterion stricter (or more lax), values other
than a half could be used.  Of course, other information about an
article, such as the subfield or author, is also potentially important
and is covered later in section \ref{sec:future}.

\section{Previous Work}

The administrative tasks associated with running an electronic journal
have already been automated by the Berkeley Electronic Press [1].  The
associated software supports the traditional peer review model and is
not freely available as open source code.

There has also been previous work specifically on automatically rating
reviewers for academic journals using collaborative filtering
techniques, for example see [2].  There is also a close connection
with work that uses collaborative filtering techniques to suggest
movies, books, music, etc. [3].  Typically, reviews are directly rated
by their degree of ``helpfulness'' to the user.  In contrast, the
readers of the \jndm only see the results of the reviewing process,
never the individual reviews themselves.

In addition, the goal of collaborative filtering approaches is usually
to find new examples that match a person's subjective tastes.  There
are, however, applications where subjective taste is not appropriate.
For example, articles that appear in a scientific journal are supposed
to express more than the subjective taste of the editorial board and
reviewers.  In particular, they are meant to embody some notion of
objective quality.  If they do not, then the scientific theories they
espouse will founder upon their first encounter with reality.  Of
course, within the wide selection of high-quality academic articles
that are available, there is still an important place for an
individual's personal preferences and interest.

The goal of the \jndm is not to help readers find articles that they
are interested in, but to use reader's feedback to produce better
quality publications.  Whether this really amounts to an important
distinction is clearly open to question.  The argument is that the
democratic nature of the reviewing process allows input from a wider
group of people and is thus more likely (on average and over time) to
produce better quality results.  The \jndm exists to see if this
hypothesis is at least plausible.  Even if the \jndm turns out to be a
failure, the reasons for failure may turn out to have interesting
ramifications for other democratic processes.

Another key difference between this paper and previous work is that it
describes a working prototype, namely the \jndmp.  Trying to actually
run a democratic journal should yield important insights into the
associated challenges and possible solutions.  

\section{Submitting an Article}

The process of submitting an article to the \jndm is straightforward
and should be familiar to anyone who has electronically submitted an
article to any other journal or conference.

The author creates an account at the \jndm web site
(\url{www.democraticjournals.org})
and then submits her article.  Currently, the
article submission process leverages the existing e-Print archive at
\url{arXiv.org}.  That is, the author submits a link to her article
that she previously uploaded to arXiv.  The reason for this is
twofold:
\begin{enumerate}
\item Until the \jndm website can build up enough history
to automatically detect articles that qualify as ``spam'', the arXiv
site provides some minimum barrier to entry.
\item The current \jndm website has limited storage available to
store large files.
\end{enumerate}

Once an article is submitted to the \jndmp, relevant information about
the submission is stored in a database.  After a fixed time period a
decision is made, based on the received reviews, on whether to accept
or reject a submission.

\section{Reviewing an Article}

Anyone can register at the \jndm web site and any registered user (not
just authors) can retrieve a list of submissions that need to be
reviewed.  Some submissions will be closer to a decision being taken
on whether to accept or reject and these appear at the top of the
list.  If an article does not receive enough (currently two) reviews
in time, then it is rejected without prejudice and can be resubmitted.
 
When a user selects an article to review they are asked to recommend
acceptance or rejection.  In future, perhaps when more data is
available, additional levels of gradation could be added but for now a
binary decision is simplest.  In addition, there is currently no
mechanism to provide feedback to authors on why their submission was
rejected or accepted.  In future, feedback from reviewers who were
influential in the decision to accept or reject a submission should be
made available to authors.

\begin{figure}[htbp]
\centerline{\epsfig{figure=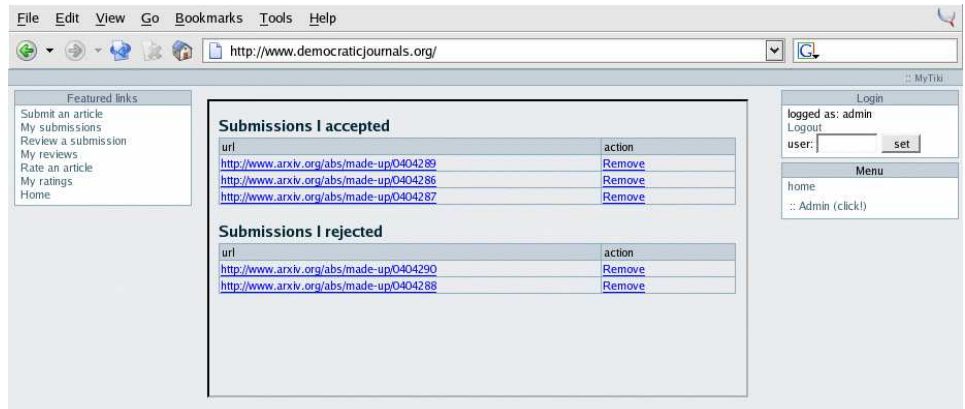,width=\textwidth,silent=}}
\caption{Screen shot from an example page of the \jndmp.}
\label{fig:screen}
\end{figure}

Review results are also stored in a database and a reviewer can view
their own previous reviews at any point.  Provided the journal has not
made a final decision on whether to accept or reject an article, a
reviewer is free to alter or delete their own past reviews.  Figure
\ref{fig:screen} shows a screen shot from an example page of the
\jndmp.

If the \jndm becomes popular, additional security auditing will be
required to ensure there is no way for unscrupulous users to obtain or
alter restricted information contained in the journal's database.

\section{Reading and Rating an Article}
\label{sec:read}

Any registered user can read and rate any published article to express
whether or not they feel the decision to publish an article was
acceptable.  Once again, the rating is (for now) a binary decision.
Just like reviewers, readers can manage their past ratings and alter
them if they change their minds or make a mistake.

Currently, there is no special action taken when a reader who was also
a reviewer for an article expresses an opinion that is inconsistent
with their review, for example, if a reader states an article is
unacceptable that they previously recommended to accept.  In future,
inconsistencies should at least be flagged to the user to ensure they
really did mean to change their mind.

\section{Publication}
\label{sec:publication}

As stated in the Introduction, an article is published in the \jndm if
the conditional probability that it will be considered acceptable by a
majority of the readers is, given the reviews, above a half.

Since each review is (at least, supposed to be) independent, and
because it is one of the simplest thing to try first, Naive Bayes (see
[4]) is currently used to calculate the probability that an article is
acceptable.  Note that, this paper therefore presents no new machine
learning algorithms, just a novel application of known techniques.
But, for the sake of completeness, the derivation of the required
formula follows using the standard application of Bayes rule and the
assumption of independent reviews:
\begin{eqnarray}
P( \text{acceptable}~|~\vec{r} ) & = & \frac{P( \vec{r}~|~\text{acceptable} ) P( \text{acceptable} )}{P( \vec{r} )} \nonumber \\
                                 & = & \frac{P( r_1~|~\text{acceptable} ) \cdots P( r_m~|~\text{acceptable} ) P( \text{acceptable} )}{P( \vec{r} )},
\label{eqn:predict}
\end{eqnarray}
where the conditional probabilities $P( r_i~|~\text{acceptable} )$ and
the prior $P( \text{acceptable} )$ are determined from historical data
(see the next section).  As usual, the final probability is computed
by also calculating $P(\text{unacceptable}~|~\vec{r} )$ and
normalizing.

Naive Bayes was a reasonable choice to quickly bootstrap the \jndm
into existence and is a good benchmark to beat, but there are, of
course, many other machine learning techniques that could (and
probably should) be used instead.  For example, decision trees [4] and
support vector machines [5] are both good candidates.  You are
strongly encouraged to submit an article to the
\jndm containing your own opinions, objections, observations,
improvements and alternatives.  Ideas on algorithms and techniques to
make the journal fairer and more resilient to exploitation are
particularly encouraged.

\section{Reviewers}

As stated in the Section \ref{sec:read}, after an article is accepted
and published in the \jndmp, people who read the article can express
their opinion about whether the article deserved to have been
published.  Whenever a new publication decision must be taken, the
opinions of the readers of each article are tallied and, if the number
of positive votes is greater than the negative votes, the article is
labeled as acceptable.  The label of acceptable, or unacceptable, is
used to create the training set from which the probabilities $P(
r_i~|~\text{acceptable} )$ and $P( r_i~|~\text{unacceptable} )$ are
calculated using frequency counts.  The prior $P( \text{acceptable} )$
is just taken as the journal's overall current acceptance rate.

\begin{table}[htbp]
\begin{center}
\begin{tabular}{c|c|c|c|c|c}
           & $R_1$ & $R_2$ & $R_3$ & $R_4$ & Acceptable? \\ \hline
Article 1  &   1   &   1   &   0   &   0   &   1 \\
Article 2  &  --   &   0   &  --   &   0   &   1 \\
Article 3  &  --   &   1   &   0   &   1   &   0 \\
Submission &   0   &   1   &   0   &   1   &   ?
\end{tabular}
\end{center}
\caption{Simple example of historical data.}
\label{tab:data}
\end{table}

In case you do not have a statistics or machine learning background,
Table \ref{tab:data} shows a simple pedagogical example containing
some made-up historical data.  Articles 1 through 3 are articles that
have previously been accepted or rejected and the ``Submission'' is
the article that is currently being considered for publication.  The
reviewers $R_1$ through $R_4$ are the reviewers who submitted (for the
current submission) reviews $r_1$ through $r_4$.  The last line shows
the results of those reviews (i.e., $\vec{r} = ( 0, 1, 0, 1 )$), and
the previous lines show the reviewer's decisions (if any) on previous
publications.

From the table you can see that $P( r_2~|~\text{acceptable} )$, for
example, is $\frac{1}{2}$ and $P( r_2~|~\text{unacceptable} )$ is $1$.
In practice, a \emph{Laplace estimator} is used in place of the
frequency counts.  Otherwise, especially when there is not much data,
one of the individual probabilities can easily collapse to $0$ which
causes the whole multiplication in equation \ref{eqn:predict} to
collapse to $0$.  For example, the frequency count gives $P(
r_4~|~\text{acceptable} ) = 0$ whereas the Laplace estimator gives a
probability of $\frac{1}{3}$.  In future, Lidstone's estimator may be
used instead of the Laplace estimator (see [6] if you want to be
reminded about the Laplace and Lidstone estimators).

Currently, reviews are excluded from participating in the publication
decision until the associated reviewer has recorded at least two
previous reviews for published articles.  The articles must have been
published because (in the interests of author privacy and simplicity)
there is no label for unpublished articles as readers never get to
judge if they were correctly rejected.

\subsection{Lead Reviewers}

The \emph{precision} of a reviewer is defined using the standard
definition of precision, i.e.,
\[
\text{precision} = \frac{\text{tp}}{\text{tp} + \text{fp}},
\]
where tp is the number of true positives (articles that were correctly
accepted by the reviewer) and fp is the number of false positives
(articles that were incorrectly accepted).  More sophisticated
measures, such as the so-called \emph{F measure} [6], can not be
calculated because there is no information available about false
negatives and true negatives.

The names of the reviewers with the highest precision (based on past
performance) are published as the journal's list of lead reviewers.
Lead reviewers have the satisfaction of knowing that they have
obtained the distinction democratically.  Anyone can become a reviewer
and if you consistently act as a good representative of the journal's
wider readership you can obtain the prestige of becoming a lead
reviewer.

The open and democratic nature of the reviewing process should mean
that the \jndm can quickly adapt to new kinds of submissions, or the
changing abilities of reviewers.  To improve the rate of adaptation,
it might become necessary to weight reviewer's recent reviews more
heavily than past ones.  If so, then it would make sense to treat the
whole problem of determining the weights assigned to reviews as an
online learning problem and perhaps use techniques like those
described in [7].

\subsection{Automatic Reviewing}

Eventually, the \jndm will make statistics available about which
published articles were subsequently found acceptable by the readers.
This allows machine learning programs to compete to become reviewers.
That is, the task of predicting whether a published article will be
found acceptable is a standard supervised learning problem.  The
training set consists of the published articles labeled by whether
they were subsequently found acceptable or not.  Machine learning
algorithms that are trying to become reviewers have the difficult, but
interesting, task of trying to analyze an article's content to predict
if it should be accepted.  It will be fascinating to see how good an
indicator a feature like an article's author or affiliation are in
correctly determining whether it should be accepted.

In an attempt for even further automation, a machine learning
algorithm called the \emph{privileged reviewer} will also be
introduced.  Just like the other reviewers, the privileged reviewer
tries to predict which articles will be popular among readers.  Unlike
the other reviewers (including regular reviewers that are learning
algorithms), the privileged reviewer is given access to privileged
information about rejected submissions.  In the standard tradition of
academic journals, information about rejected submissions is not made
public.  Presumably, provided suitable privacy precautions are taken,
authors will not mind rejection information being made available to a
machine learning algorithm.  If the privileged reviewer can become one
of the lead reviewers then the need for any human input in the review
process is reduced.

The software for the privileged reviewer will also be made open source
and people will be encouraged to submit their own privileged reviewer
algorithms.  Within the confines of protecting author privacy, some
kind of test harness will be made available for researchers to
evaluate their designs for a privileged reviewer.

\section{Editors}

The editors of the \jndm are the SourceForge project administrators
for the software that runs the journal's web site.  This software
includes the code to learn which reviewers are most likely to give
good reviews, and to decide, based on the reviews, whether an article
should be published or not.  The software is expected to constantly
change (especially to start with) in response to problems that arise.
For example, if a flaw is discovered in the software that could be
exploited to make it easier to publish articles, then the editors
would be responsible for fixing the flaw.

You will have noticed by now that, in the best traditions of computer
science, the \jndm is wonderfully recursive.  It is a journal about
the very techniques that it uses to determine which articles to
publish.  Therefore, for inspiration on how to fix flaws and generally
enhance the journal, it is expected that the editors will turn to
ideas published in the \jndm itself.  The idea that articles published
in the journal feedback into the journal's software to act as a sort
of built in defense mechanism against abuse and subversion is
appealing; it will be interesting to see how it plays out in practice.

Note that, even if the editors themselves become corrupt, self-serving
or simply disliked, then (because all the journal's software is open
source and freely available) there is nothing to stop a new group of
people using the journal's software to establish their own journal
with new editors who are more to their liking.  The open source nature
of the code behind the \jndm is therefore an important component
(along with the machine learning) in making it more democratic.

\section{Future Work}
\label{sec:future}

There is currently no restriction on an author reviewing their own
articles.  Of course, if they consistently give high reviews to their
own work which later turns out to be unpopular, then they will soon
loose their influence over the review process.  Another alternative
would be to create two pseudo-reviewers for each reviewer, one for
reviewing their own work and another for reviewing the work of others.
That way an unscrupulous, but otherwise competent reviewer could
quickly loose all credibility for reviewing their own work, but keep
their good standing for reviewing the work of others.

In theory, the appropriate use of pseudo-reviewers could automatically
weed out other forms of bias, but it maybe easier to simply make
authors anonymous.  Pseudo-reviewers could perhaps be more useful for
different subfields within the journal.  That is, a reviewer could
become a lead reviewer in their area of expertise but still feel free
to try their hand at reviews in less familiar areas without penalizing
their existing standing.


Readership opinion is only one possible source of information about
the quality of articles published in the \jndmp.  Other possibilities
include citations in other journals and publication or related
articles in other journals.  For example, publication in a prestigious
journal of a rejected article could be used to label an article as
incorrectly rejected.

There is also the possibility of establishing a journal entirely based
upon benchmarks and challenge problems.  That is, articles are only
published if authors submit programs that perform well on some
objective and freely available set of tests.  This idea is already
used informally by funding agencies and academic conferences who hold
competitions to rate different research projects.


\section{Conclusion}

One of the biggest challenge to the success of the \jndm is obtaining
participation from qualified people.  This is a classic ``chicken and
egg'' problem and hopefully the paper you are now reading will help to
interest people.  The more widely it can read the better.

Even if people begin signing up to the \jndm and submitting articles
it will still be a challenge to encourage people to review and rate
articles.  It may be necessary to institute incentives to encourage
people to actively participate.  For example, users should probably
not be allowed to read another submission until they have submitted
their review for the last one.  Similarly readers could be rationed to
only being able to read a fixed number of articles without submitting
any ratings.  The danger is that a small number of enthusiasts end up
doing all the reviewing with little or no feedback from the larger
readership.

As the title suggests, the \jndm is not confined to articles about
academic journals.  An academic journal is just one example of an
institution where objective quality is important and a small group of
people try to anticipate and guide the interests of a much larger
group.  Another interesting example is the whole notion of
representative government and democracy.  Finding a small group of
individuals who can faithfully act as representatives for the people
as a whole, and be successful at running the country, is what
representative government is all about.  The use of machine learning
techniques as described in this paper, and (hopefully) to be described
in forthcoming articles in the \jndmp, could therefore have a profound
effect on society at large.


\subsubsection*{Acknowledgments}

Thanks to Ian Wright for many useful and inspiring discussions on the
topic of new democratic methods.  The web site for the \jndm is
implemented in PHP [8] using many open source tools, notably the resources
of SourceForge [9] (including a MySQL database [10]), and the Tiki
CMS/Groupware (aka TikiWiki) open source content management system [11].
Without all of these tools developing the \jndm web site in such a
short space of time would not have been possible.

\subsubsection*{References}

\small{

[1] \emph{The Berkeley Electronic Press}.
\url{www.bepress.com/revolution.html}

[2] Tracy Riggs and Robert Wilensky. (2001)
\emph{An Algorithm for Automated Rating of Reviewers}.
In Proceedings of the first ACM/IEEE-CS joint conference on Digital libraries.
\url{portal.acm.org/citation.cfm?id=379731}

[3] \emph{Collaborative Filtering}.
\url{pespmc1.vub.ac.be/COLLFILT.html}

[4] Tom M. Mitchell (1997).
\emph{Machine Learning}. McGraw-Hill, {NY}.
\url{www-2.cs.cmu.edu/~tom/mlbook.html}

[5] Vladimir Vapnik. (1999).
\emph{The Nature of Statistical Learning Theory} (Second Edition). Springer, NY.
\url{www.kernel-machines.org}

[6] Chris Manning and Hinrich Sch\"{u}tze. (1999).
\emph{Foundations of Statistical Natural Language Processing}.  MIT Press, MA.
\url{www-nlp.stanford.edu/fsnlp/}

[7] Avrim Blum. (1997) Empirical Support for Winnow and
\emph{Weighted-Majority Algorithms: Results on a Calendar Scheduling Domain}.
In Proceedings of 12th International Conference on Machine Learning.
\url{citeseer.ist.psu.edu/blum97empirical.html}

[8] \emph{PHP: Hypertext Preprocessor}.
\url{www.php.net}

[9] \emph{Open Source software development website}.
\url{sourceforge.net/}

[10] \emph{MySQL: the world's most popular open source database}.
\url{www.mysql.com}

[11] \emph{Tiki CMS/Groupware open-source Content Management System (CMS)}.
\url{tikiwiki.org}
}

\end{document}